\documentclass[prd,superscriptaddress,amsfonts,amssymb,amsmath,showpacs,twocolumn]{revtex4-2}
\usepackage{bm}
\usepackage{amsfonts}
\usepackage{latexsym}
\usepackage{graphicx}
\usepackage{amsmath}
\usepackage{palatino}
\usepackage{mathpazo}
\usepackage{textcomp}
\usepackage{float}
\usepackage{booktabs}
\usepackage{dcolumn}
\usepackage{booktabs}
\usepackage{multirow}
\usepackage{hyperref}
\hypersetup{colorlinks,citecolor=blue}
\usepackage{amsmath}
\usepackage{xcolor}
\usepackage{orcidlink}
\usepackage{epsfig}
\usepackage{caption}
\usepackage{subcaption}
\usepackage{commath}
\hypersetup{colorlinks,citecolor=blue}
\hypersetup{colorlinks=true,linkcolor=magenta,filecolor=magenta,    urlcolor=blue}

\def\be{\begin{equation}}
\def\ee{\end{equation}}
\def\bea{\begin{eqnarray}}
\def\eea{\end{eqnarray}}

\begin{document}

\title{Dark Sector Interactions: Probing the Hubble Parameter and the Sound Horizon}

\author{Ritika Nagpal}
\email{ritikanagpal.math@gmail.com,\\
ritikanagpal@vivekanand.du.ac.in}
\affiliation{Department of Mathematics, Vivekananda College, University of Delhi, New Delhi, India}
\author{S. K. J. Pacif}
\email{shibesh.math@gmail.com}
\affiliation{Pacif Institute of Cosmology and Selfology (PICS), Sagara, Sambalpur 768224, Odisha, India} 
\affiliation{Research Center of Astrophysics and Cosmology, Khazar University, Baku, AZ1096, 41 Mehseti Street, Azerbaijan}
\author{Farruh Atamurotov}
\email{atamurotov@yahoo.com}
\affiliation{University of Tashkent for Applied Sciences, Str. Gavhar 1, Tashkent 100149, Uzbekistan}
\affiliation{Urgench State University, Kh. Alimdjan str. 14, Urgench 220100, Uzbekistan}
\affiliation{Shahrisabz State Pedagogical Institute, Shahrisabz Str. 10, Shahrisabz 181301, Uzbekistan}
\author{Rasmikanta Pati}
\email{rkpati@suiit.ac.in} 
\affiliation{SUIIT, Sambalpur University, Sambalpur 768019, Odisha, India}

\begin{abstract}
In this study, we explore the impact of the interacting parameter on dark matter in a model resulting from a parametrization of dark energy density. To ensure a model-independent approach, we treat \( r_d \) as a free parameter, avoiding assumptions about the physics of the early Universe or specific recombination models. This approach allows late-time cosmological observations to directly constrain \( r_d \) along with other parameters. Using recent measurements from the Dark Energy Spectroscopic Instrument (DESI) Year 1, cosmic chronometers (CC) and Pantheon\(^{+}\) supernova (SNe Ia) data, we uncover a significant effect of the interacting parameter on dark matter. Our analysis reveals that while non-interacting models attribute 68.2\% of the cosmic energy density to dark energy, interacting models increase this share to 73.4\%. To further probe these differences, we evaluate the evolution of the deceleration parameter for each model, contrasting them against the \(\Lambda\)CDM paradigm and observational data from CC and SNe Ia measurements. Finally, we apply various statistical metrics to rigorously assess the performance of these models.  
\end{abstract}

\maketitle


\section{Introduction}\label{sec1}
The late-time acceleration (LTA) of the Universe is one of the most profound mysteries in contemporary astrophysics and cosmology \cite{riess1998observational}. Since the late 1990s, when observations of distant Type Ia supernovae (SNeIa) first revealed an unexpected acceleration, a growing body of evidence from Cosmic Microwave Background (CMB) \cite{perlmutter1999measurements}, large-scale structure (LSS) surveys, and baryon acoustic oscillations (BAO)—has consistently affirmed this remarkable phenomenon \cite{tegmark2004cosmological,seljak2005cosmological,eisenstein2005detection}. Additional insights from weak gravitational lensing \cite{jain2003cross}, as well as results from WMAP \cite{hinshaw2013nine}, the Sloan Digital Sky Survey (SDSS) \cite{anderson2014clustering}, and Planck collaborations, have further strengthened the case for LTA \cite{aghanim2020planck,aghanim2020planck1}. Understanding the underlying cause of this acceleration presents a formidable challenge. Broadly, two theoretical avenues have been pursued. The first involves introducing new components into the cosmic energy budget—most notably dark energy, an enigmatic form of energy with negative pressure capable of driving repulsive gravitational effects. The second pathway considers modifying the geometric underpinnings of General Relativity (GR) itself, for instance by altering the Einstein Hilbert action to accommodate deviations from standard gravity. Both approaches reflect our ongoing effort to reconcile the observed accelerating expansion with a coherent and predictive theoretical framework. In doing so, they highlight the need for continued observational precision and innovative theoretical models to reveal the true nature of LTA.\\\\
The cosmological constant \(\Lambda\) has long served as the canonical explanation for dark energy (DE), offering a simple and observationally concordant model of late-time cosmic acceleration. Nonetheless, the so-called “cosmological constant problem” — the vast discrepancy between theoretical predictions of vacuum energy and its observed value—continues to undermine this elegant solution \cite{weinberg1989cosmological}. In response, a spectrum of alternative DE frameworks has emerged, either invoking symmetries to suppress vacuum energy or introducing time-dependent \(\Lambda\), thereby allowing for dynamic interactions that circumvent independent conservation of dark matter (DM) and DE. Such modifications not only challenge the conventional \(\Lambda\)CDM paradigm but also tackle the related “cosmic coincidence” problem. Interacting DM-DE models, for instance, propose energy or momentum exchange between these components, effectively synchronizing their evolution over cosmic time \cite{ref1, ref2, ref3, ref4, ref5, ref6}. In contrast to scalar field models that depend on specific potentials and kinetic terms \cite{ref7, ref8}, these interacting scenarios foreground coupling mechanisms to reconcile theoretical expectations with observational data \cite{ref9, ref10, ref11, ref12}. While the \(\Lambda\)-term (\(\omega_\Lambda = -1\)) remains remarkably successful in fitting current observations, these emerging avenues enrich our theoretical toolbox and offer promising strategies to resolve enduring tensions in our understanding of the accelerating expansion of the Universe.\\\\
To retain the framework of General Relativity (GR) while accounting for the observed acceleration of the Universe, it becomes necessary to introduce a component exhibiting negative pressure. Einstein’s cosmological constant \(\Lambda\) is a principal candidate for this dark energy (DE), behaving like a fluid with equation of state \( p_{\Lambda} = -\rho_{\Lambda} \) and often associated with quantum vacuum energy. However, the substantial discrepancy between theoretical predictions and observed values of \(\Lambda\) has driven the development of alternative DE models. These frameworks may employ symmetries to suppress vacuum energy contributions or permit a time-varying \(\Lambda\), suggesting possible interactions between dark matter (DM) and vacuum energy that prevent their independent conservation. Many such constructions are formulated to recover the \(\Lambda\text{CDM}\) limit, characterized by \(\omega_{\Lambda} = -1\), which remains consistent with current observations.\\\\
The persistent fine-tuning and cosmic coincidence problems further motivate the investigation of non-standard DE paradigms, including interacting DM-DE scenarios. These posit an exchange of energy or momentum between DM and DE, thereby linking their density evolutions and mitigating the coincidence issue. In contrast to scalar field models, which depend on specific potentials and kinetic terms, interacting frameworks emphasize coupling mechanisms to reconcile theoretical predictions with empirical data. By broadening the theoretical landscape, these models offer new avenues to address outstanding tensions and deepen our understanding of the accelerating expansion of the Universe.
\\\\
The reconstruction method, originally introduced by Starobinsky for dark energy (DE) studies \cite{starobinsky1998determine}, remains a pivotal tool for probing the Universe’s expansion history \cite{turner2002type, ellis1991exact}. It is generally classified into two categories: parametric and non-parametric approaches. Parametric reconstruction employs predefined parameterizations of cosmological quantities, subsequently constrained by observational data. While successful in capturing the transition from early deceleration to late-time acceleration, its reliability depends on the chosen parametrization \cite{gong2007reconstruction, santos2011current}. Non-parametric reconstruction avoids this assumption but faces its own interpretational challenges \cite{crittenden2012fables,holsclaw2011nonparametric}. As a result, no universally accepted theory has yet emerged.\\\\
Among these methods, the parametric approach has proven particularly effective in elucidating cosmic phase transitions \cite{cunha2008transition, aich2023interacting, mukhopadhyay2024inferring,thermo2022dark, debnath2024dark,koussour2023hubble,akarsu2014probing}. Recent advancements emphasize modeling DE within the framework of General Relativity (GR) via the equation of state parameter \(\omega\), yielding results consistent with diverse observational datasets \cite{linder2003exploring, nojiri2015singular,wang2001measuring}. Notably, parametrizing DE energy density has outperformed other cosmological parameters in constraining the Universe evolution of the Universe \cite{maor2002measuring}. This study adopts a parametric reconstruction strategy centered on DE energy density, aiming to refine our understanding of cosmic phenomena and phase transitions through contemporary observational data.\\\\
The work of this paper has been organized as follows: The Section ~\ref{sec1} is introductory briefing formal presentation on dark energy and a brief proposal about the late time cosmic acceleration. In Section ~\ref{sec2}, we have discussed the Einstien field equations and obtain its GR  solutions in a model-independent way. In Section ~\ref{sec3}, we constrain the model parameters for the detailed analysis on the behavior of physical parameters using some observational datasets. We have discussed some statistical metrics in the Section ~\ref{sec4}. Also, we have explained the results in Section ~\ref{sec5}. Finally, we conclude our discussion in Section ~\ref{sec6}.
\section{Einstein's Field Equations in FLRW Spacetime and its solution}\label{sec2}
We consider a homogeneous, isotropic, and spatially flat spacetime for our geometrical framework. In the context of a flat FLRW universe, we now consider a cosmological model in which dark energy (DE) does not evolve in isolation but interacts dynamically with the cold dark matter (CDM) component, which includes both baryonic and non-baryonic matter. Such interactions can alter the standard evolution of cosmic constituents, potentially alleviating tension between observations and theoretical predictions, and providing a natural explanation for the apparent coincidence that the densities of matter and dark energy are of the same order today. By allowing for energy exchange between CDM and DE, these interacting models offer a richer phenomenological landscape, enabling a more flexible reconstruction of the expansion history and improved fits to current observational data.\\\\
In this interacting framework, the total energy-momentum tensor takes the form
\begin{equation}
T^{t}_{ij} = (\rho_{t} + p_{t})u_i u_j + p_{t}g_{ij}, \label{1}
\end{equation}
where \(\rho_{t}\) and \(p_{t}\) denote the total energy density and pressure, respectively, encompassing all cosmic components. In this framework, \(\rho_{t}\) and \(p_{t}\) represent the total energy density and pressure contributed by all cosmic components. Specifically, the total energy density is given by \(\rho_{t} = \rho_{CDM} + \rho_{de}\), where \(\rho_{CDM}\) and \(\rho_{de}\) denote the energy densities of cold dark matter (CDM) and dark energy (DE), respectively. Since CDM is effectively pressureless, the total pressure reduces to that of the DE component, \(p_{t} = p_{de}\).
The corresponding equation of state (EoS) is given by
\begin{equation}
p_{t} = w_{t}\rho_{t}, \label{2}
\end{equation}
with the parameter \(w_{t}\) generally evolving over time, thereby accommodating a more dynamic cosmic evolution than traditional, non-interacting models. 
In scenarios where cold dark matter (CDM) and dark energy (DE) actively interact, with \(\rho_m = \rho_b + \rho_{DM}\) representing baryonic and dark matter densities, and \(\rho_{DE}\), various DE candidates emerge from distinct scalar field (\(de\)) potentials. Among them, the cosmological constant \(\Lambda\), characterized by \(w_{de} = -1\), remains the simplest and most empirically supported option. This potential-energy-dominated framework integrates smoothly into interacting models, maintaining consistency with current observations.
The field equations then become
\[
8\pi G \rho_{t} = \frac{3\dot{a}^2}{a^2}, \quad
8\pi G p_{t} = -2\frac{\ddot{a}}{a} - \frac{\dot{a}^2}{a^2},
\] where $a(t)$ being the scale factor of the Universe, and an overhead dot signifies the time derivatives. These equations establishing a direct connection between the dynamics of DE and the observed accelerated expansion of the Universe.
\subsection{The Interacting Scenario}
In contrast to non-interacting cosmological models, introducing interaction terms between \(CDM\) and \(DE\) offers a compelling pathway to address fundamental challenges in modern cosmology. By coupling the evolution of energy densities to the Hubble parameter \(H\), these models naturally tackle the cosmic coincidence problem, establishing a dynamical equilibrium that synchronizes \(DE\) and \(CDM\) abundances over cosmic time. The resulting time-dependent equation of state (EoS) for \(DE\) enriches the expansion dynamics and can improve compatibility with observational datasets, including those from Planck, supernovae, and baryon acoustic oscillations. Such interactions not only allow for the formation of more diverse and realistic large-scale structures but also foster a deeper understanding of DE’s nature. In this way, interaction models serve as a flexible and robust framework that transcends the limitations of non-interacting scenarios, providing refined insights into the Universe’s accelerated expansion and its underlying physical principles. \cite{planck2018results,amendola2000coupled,gavela2010dark,wang2016dark,di2020hubble,zimdahl2001interacting,nojiri2017modified}. The Einstein Field Equations in the background of a flat FLRW space time can be rewritten as
\begin{equation}\label{3}
\Big(\frac{\dot{a}}{a}\Big)^2=\frac{1}{3} (\rho_{t})=\frac{1}{3}(\rho_m+\rho_{de}),  
\end{equation}%
\begin{equation}\label{4}
\Big(\frac{\dot{a}}{a}\Big)^2+2\Big(\frac{\ddot{a}}{a}\Big)=-p_{t}=-p_{de}, 
\end{equation}
in the units of $ 8\pi G=c=1 $.\\\\
In this work, we consider a cosmological framework in which DE and CDM do not evolve as separate, self-contained components. Instead, we allow for a non-negligible exchange of energy between them, introducing an interaction term that modifies their usual conservation laws. The CDM-DE interacting system is subject to an additional source term \(Q\). This function governs the direction and rate of energy transfer: a positive \(Q\) channels energy from DE into CDM, while a negative \(Q\) signifies the opposite flow.\\\\
Such interacting models offer a richer phenomenology, as they can potentially address persistent tensions in cosmological data and provide insights into why the energy densities of DE and CDM appear finely balanced at late times. Although the precise microphysical mechanism underlying this coupling remains uncertain, a variety of plausible forms for \(Q\) have been proposed. These range from simple, phenomenological parametrization to those inspired by underlying field-theoretic or modified gravity frameworks. In this paper, we present an alternative set of interaction terms and explore their implications for the evolution of cosmic expansion and structure formation. By analyzing how these interactions influence key observables, we aim to discern whether certain choices of \(Q\) can alleviate existing discrepancies and enhance our understanding of the subtle interplay between the dark components of the Universe.\\\\
When DE is attributed to a cosmological constant \(\Lambda\), any non-trivial exchange of energy with DM must accommodate the fixed nature of vacuum energy. Such considerations impose stringent conditions on the interaction term \(Q\), restricting the range of viable models. By contrast, if DE arises from a dynamic field—like quintessence or k-essence—then the situation becomes more intricate. In that case, the scalar field’s temporal evolution allows for a wider range of interaction behaviors, potentially complicating the relationship between DE and DM.\\\\
This inherent complexity underscores a fundamental challenge: the uncertain identity and properties of DE preclude a definitive choice of interaction form \(Q\). 
A frequently investigated approach to modeling \(CDM-DE\) interactions involves coupling the energy exchange rate to the cosmic expansion and the energy densities of the dark components. One strategy, for instance, posits a coupling parameter \(\eta\) and expresses the interaction as
\begin{equation}
Q = \eta H \rho_{dm} \label{5}
\end{equation}
where \(H\) is the Hubble parameter. By relating \(Q\) to \(H\), the interaction strength naturally evolves with the expansion history of the Universe. In this manner, the parameter \(\eta\), encapsulates the coupling strength, allowing the interaction to vary in step with changing cosmological conditions and providing a flexible framework for examining how energy transfer between the dark components affects cosmic dynamics.\\\\
In the interacting framework described earlier, the conservation equations can be split into separate balance equations for each component. For non-relativistic baryonic matter, solving the continuity equation under standard assumptions leads to
\begin{equation}\label{6}
\rho_b = \frac{B}{a^{3}},
\end{equation}
where \(B\) is an integration constant, and \(\rho_b\) evolves as \(a^{-3}\), reflecting the expected dilution of baryonic matter with the Universe's expansion.\\\\
We now specify the interaction term as
\begin{equation}\label{7}
Q = 3\gamma H \rho_{dm}, 
\end{equation}
where \(3 \gamma= \eta\) is a dimensionless coupling parameter and \(\rho_{dm}\) denotes the dark matter density. Substituting this form of \(Q\) into the modified conservation equations for the \(CDM-DE\) interacting system, and integrating, yields
\begin{equation}\label{8}
\rho_{dm} = \frac{D}{a^{3(\gamma - 1)}},
\end{equation}
where \(D\) is another integration constant. The corresponding equation for the dark energy density \(\rho_{de}\) becomes
\begin{equation}
\dot{\rho}_{de} + 3\frac{\dot{a}}{a}\left(1 + \omega_{de}\right)\rho_{de} = -3H\gamma D\,a^{3(\gamma - 1)}, \label{9}
\end{equation}
indicating that the DE density evolution is explicitly coupled to the \(CDM\) component via the interaction parameter \(\gamma\). In this interacting scenario, the dark energy equation of state parameter \(\omega_{de} = \frac{p_{de}}{\rho_{de}}\) serves as a fundamental characteristic that distinguishes and classifies various dark energy models.
\subsection{The Model}
In our earlier work \cite{singh2020flrw}, where we had introduced the \textit{Energy Density Scalar Field Differential (EDSFD)} parametrization in a non-interacting cosmological framework. The result we had found were compelling to the current scenario of dark energy modeling. Here, in this study, we aim to extend that approach to an interacting scenario. The core motivation for this extension lies in addressing the dynamics of energy exchange between DE and CDM, which has gained significant attention in recent cosmological research\cite{giare2024interacting,li2024constraints,rugg2024interacting,van2024interacting}.\\\\
In \cite{singh2020flrw}, we parametrized the scalar field energy density ($\rho_\phi$) through a differential equation dependent on the scale factor, $a$. This approach effectively captured the phase transition from early deceleration to present acceleration by reconstructing the equation of state parameter ($\omega_\phi(z)$) in terms of redshift. The differential equation form not only ensured mathematical consistency but also provided flexibility to model the evolution of dark energy within the FLRW framework.\\\\
However, in our study (interacting scenario), where DE and CDM exchange energy, the continuity equation (\ref{9}) which inherently appears as a differential equation requires special attention. The evolution of the energy densities in this case depends critically on the choice of a functional form $f(a)$, which influences the nature of solutions. Given that the coefficient terms on the LHS of equation (\ref{9}) depends explicitly on the scale factor, selecting an appropriate function $f(a)$ becomes a fundamental aspect of solving the continuity equation and exploring the interaction dynamics.\\\\
Thus, this work aims to generalize the earlier non-interacting framework \cite{singh2020flrw} by introducing an interacting term in the continuity equation. By carefully formulating the function $f(a)$, we seek to model the energy transfer mechanism and investigate its impact on the evolution of the Universe, particularly its transition phases. This extension not only strengthens the foundation laid by the earlier parametrization but also opens pathways to study broader scenarios, offering deeper insights into the dynamics of cosmic acceleration driven by interacting dark energy models.\\\\
With the above set up, we have to solve the system of governing equations for which we need to consider one more constraint for a deterministic solution. There are several ways to choose a constraining equations and is discussed comprehensively in \cite{pacif2020dark}. Here, we have a different approach as mentioned in the above paragraph. Following \cite{singh2020flrw}, we consider a liner differential equation in energy density of dark energy in the form
\begin{equation}
\rho_{de}'+\beta f(a)\rho_{de}=0,\label{10}
\end{equation}
where '$'$' represents differentiation w.r.t. '$a$' and $f(a)$ is an arbitrary function. The choice of the function provide the unique behavior of dark energy density. In our previous work \cite{singh2020flrw}, we have chosen the function \(f(a)\) to induce a signature-flip for the deceleration parameter \(q\). This strategic choice ensures a natural transition from an early deceleration era—suitable for structure formation—to a regime of late-time cosmic acceleration, aligning with current observational evidence. Encouraged by the versatility of this approach, we now turn our attention to examining the evolution of \(\rho_{de}\) within an interacting scenario. By exploring how energy exchanges between CDM and DE might influence the behavior of \(\rho_{de}\), we aim to address persistent theoretical challenges, such as the cosmic coincidence problem, within the framework of General Relativity. This investigation promises to shed new light on the interplay between dark components and the fundamental properties of our accelerating Universe.
\subsection{Significance of Parametrizing the Energy Density of Dark Energy}
\begin{enumerate}
    \item \textbf{Direct Physical Influence on Cosmic Expansion} Parametrizing the DE density \(\rho_{\text{DE}}\) enables a direct link to the Friedmann equations, allowing precise examination of how changes in \(\rho_{\text{DE}}\) drive the expansion of the Universe and influence parameters like the Hubble parameter \(H\) and the deceleration parameter \(q\) \cite{huterer1999prospects, wang2004model}.
    \item \textbf{Improved Constraints from Observational Data} Employing \(\rho_{\text{DE}}\) parameterizations provides a more flexible framework for fitting observational data, including supernovae, baryon acoustic oscillations (BAO), and the cosmic microwave background (CMB), thereby yielding tighter constraints on DE properties and revealing subtle temporal variations \cite{maor2001dynamics, chevallier2001accelerating}.
    \item \textbf{Modeling Complex, Time-Dependent Dynamics} By treating \(\rho_{\text{DE}}\) as a time-varying quantity, one can capture complex evolutionary behaviors of DE that address conceptual challenges such as the cosmic coincidence problem and accommodate a broad range of dynamical scenarios \cite{linder2003exploring}.
    \item \textbf{Bridge Between Theory and Data} Many theoretical models predict specific functional forms for \(\rho_{\text{DE}}\). Parametrizing this quantity creates a pathway for testing these predictions against observations, refining or refuting theoretical frameworks and advancing the overall understanding of DE \cite{wang2004model}.
    \item \textbf{Insights into Phase Transitions}
    Understanding how \(\rho_{\text{DE}}\) evolves over cosmic time is critical for identifying transitions, such as the shift from matter domination to DE domination, and for pinpointing when and how these changes occur \cite{maor2001dynamics, caldwell2003phantom}.
    \item \textbf{Unified Interpretation of Diverse Observations} A common parameterization of \(\rho_{\text{DE}}\) supports the synthesis of multiple datasets within a single framework, helping to resolve observational tensions - such as different measurements of \(H_0\) - and promoting a more coherent understanding of the expansion history of the Universe. \cite{valentino2020hubble}.
\end{enumerate}

With the above motivation, we choose an analytic function with one parameter in the form 
\[
f(a)=\beta^2\left[ 1+\frac{1}{a\sqrt{1+a^{2\beta}}\sinh^{-1}(1/a)^{\beta}} \right].
\]
Now, the general solution of the above differential equation (\ref{10}) for this function $f(a)$ is given by
\begin{equation}
\rho_{de}=e^{-\beta a} sinh^{-1}\Big(\frac{1}{a}\Big)^{\beta},\label{11}
\end{equation}
where $\beta \in (0,1)$ is the model parameter.\\\\
Utilizing the standard relation between redshift \(z\) and scale factor \(a\), we have
\[
\frac{a}{a_0} = \frac{1}{1+z}.
\]
Expressing the scalar field energy density \(\rho_{de}\) in terms of redshift \(z\),
\begin{equation}
\rho_{de}(z)=e^{ \frac{-\beta}{1+z}} sinh^{-1}(1+z)^{\beta},\label{12}
\end{equation}
and 
\begin{equation}
\rho_{de_{0}}=e^{-\beta} sinh^{-1}(1).\label{13}
\end{equation}    
Equations (\ref{12}) and (\ref{13}) yield
\begin{equation}
\rho_{de}(z)= \frac{\rho_{de_{0}}}{sinh^{-1}(1)} e^{\frac{\beta z}{1+z}} sinh^{-1}(1+z)^{\beta},\label{14}
\end{equation}
where $\rho_{de_{0}}$ is the present value of the DE energy density.
Also using equations (\ref{6}), (\ref{8}) and (\ref{14}) in equation (\ref{3}), we have
\begin{equation}
\begin{split}
 3H^2 &= B(1+z)^3+ D(1+z)^{3(1-\gamma)}+\frac{\rho_{de_{0}}}{sinh^{-1}(1)}\\ &\quad e^{\frac{\beta z}{1+z}} sinh^{-1}(1+z)^{\beta}.\label{15}
 \end{split}
\end{equation}
We define the density parameter \(\Omega=\frac{\rho}{\rho_c}\), where \(\rho_c=\frac{3H^2}{(8\pi G)^2}\) denotes the critical density. In normalized units, we set \(8\pi G=1\), simplifying the resulting expressions.\\\\
Equation (\ref{15}) in terms of density parameter of CDM and DE can be expressed as
\begin{equation}
\begin{aligned}
H(z) &= H_0 \bigg[ \Omega_{b_0}(1+z)^3 + \Omega_{dm_{0}}(1+z)^{3(1-\gamma)}\\
&\quad + \frac{\Omega_{de_{0}}}{\operatorname{arcsinh}(1)} \left( e^{\frac{\beta z}{1+z}} \operatorname{arcsinh}(1+z)^\beta \right) \bigg]^{1/2},
\end{aligned}
\label{16}
\end{equation}
where $\Omega_{b_0}=\frac{B}{3H_0^2}$, $\Omega_{dm_0}=\frac{D}{3H_0^2}$ and $\Omega_{de_0}=\frac{\rho_{de_{0}}}{3H_0^2}$ are the present values of baryonic matter, DM and DE density parameters respectively.\\\\
The deceleration parameter \(q\) provides a direct measure of the acceleration state of the Universe’s expansion. Defined as
\[
q = -\frac{\ddot{a}a}{\dot{a}^{2}} = -1 - \frac{\dot{H}}{H^{2}},
\]
it captures the sign and magnitude of the cosmic acceleration. In observational terms, the deceleration parameter can also be expressed as
\[
q(z) = -1 + (1+z)\frac{d\ln H(z)}{dz},
\]
relating it directly to the Hubble parameter \(H(z)\) and redshift \(z\). This formulation allows for the use of current and forthcoming cosmological data to probe the acceleration or deceleration behavior of the Universe more precisely.
\begin{figure}[htbp]
\centering
\includegraphics[scale=0.41]{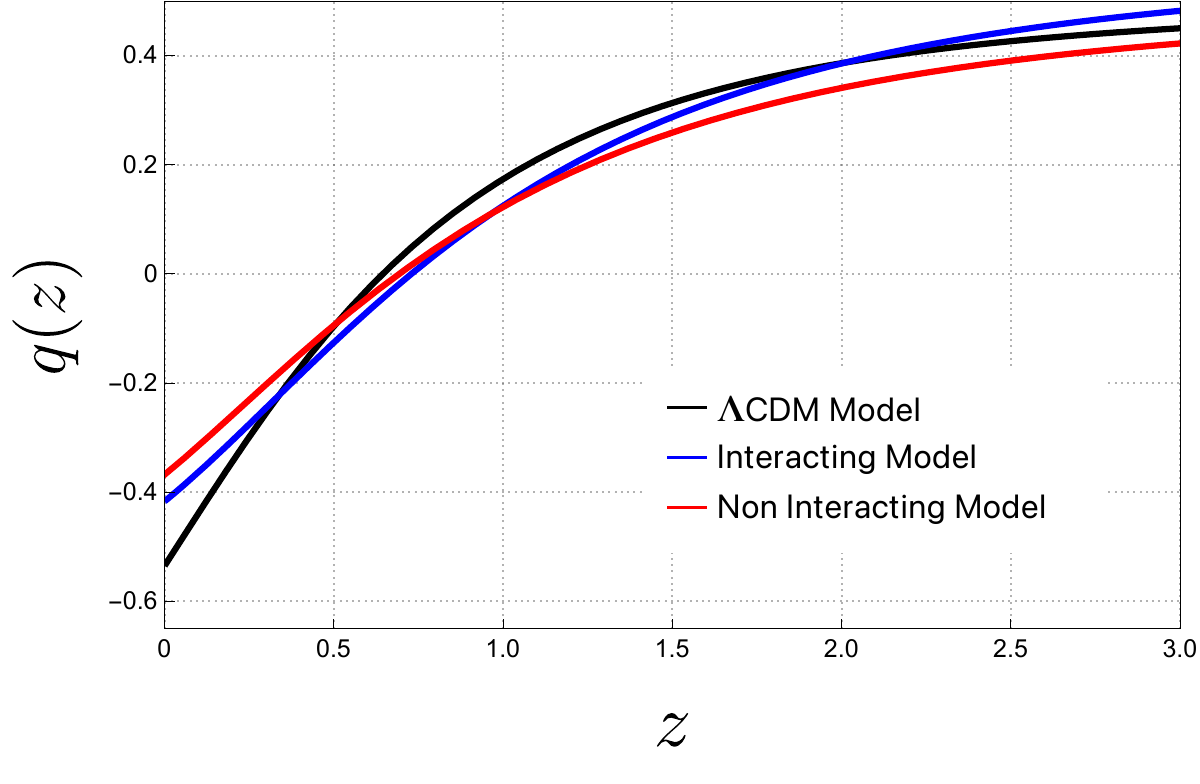}
\caption{The figure shows the evolution of deceleration parameter as a function of redshift ($z$).}\label{fig_1}
\end{figure}
As a consistency check and to gain further insight into the interacting framework, we examine the conditions under which it reduces to the standard non-interacting model. In the limiting case where the interaction term vanishes or if we set the interaction parameter $\gamma$ to zero (or let the coupling parameter approach zero), all energy transfer terms between the dark components vanish, the interacting model naturally reduces to the non-interacting scenario, thereby recovering the standard cosmological framework. In particular,  Under this limit, the evolution equations recover their familiar non-interacting forms, and the model transitions smoothly to the standard scenario where CDM and DE evolve independently. This provides a useful benchmark, allowing us to compare and contrast the properties and predictions of the interacting model with those of the well-established non-interacting paradigm.
\begin{equation}
\begin{split}
H(z) &= H_0 \bigg[ \Omega_{b_0}(1+z)^3 + \Omega_{dm_{0}}(1+z)^3 \\
     &\quad + \frac{\Omega_{de_{0}}}{\operatorname{arcsinh}(1)} \left( e^{\frac{\beta z}{1+z}} \operatorname{arcsinh}(1+z)^\beta \right) \bigg]^{1/2},
\end{split}\label{17}
\end{equation}
where \(\beta\) is the model parameter.\\\\
\section{Methodology and Data Description}\label{sec3}
To estimate the distribution of cosmological parameters for our model in both interacting and non-interacting scenarios, we apply the \texttt{PyPolyChord} nested sampling algorithm \cite{handley2015polychord}. This method utilizes Bayesian inference to calculate the posterior distribution, defined as:
\[
P(\theta | D) = \frac{L(D | \theta) P(\theta)}{P(D)},
\]
where \( P(\theta | D) \) is the posterior, \( L(D | \theta) \) is the likelihood, \( P(\theta) \) is the prior, and \( P(D) \) is the evidence. Unlike traditional MCMC methods, \texttt{PyPolyChord} uses nested sampling \cite{skilling2006nested}, refining live points iteratively to efficiently explore the parameter space and handle complex posterior distributions. We define the likelihood function using a \( \chi^2 \) statistic to assess the model's consistency with the data. Uniform priors are chosen based on prior knowledge, and a custom uniprior function maps the unit hypercube to the actual parameter space, ensuring proper exploration. The 100 live points, with clustering enabled to enhance exploration. Convergence is reached when the posterior mass within the live points reaches \( p = 10^{-2} \) of the total calculated evidence.\\\\
\begin{itemize}
    \item Cosmic Chronometers: We consider the compilation of 31 measurements of the Hubble parameter \( H(z) \) over the redshift range \( 0.07 < z < 1.965 \), as summarized in Table 1 of \cite{vagnozzi2021eppur}. The chi-square statistic for this dataset is given by the following expression: $\chi^2_{\text{CC}} = \Delta \mathbf{H}^T(z) \mathbf{C}^{-1} \Delta \mathbf{H}(z),$ where \( \Delta \mathbf{H}(z) = \mathbf{H}_{\text{model}}(\theta) - \mathbf{H}_{\text{obs}} \) represents the difference between the theoretical Hubble parameter, \( \mathbf{H}_{\text{model}}(\theta) \), at redshift \( z_i \) (calculated for model parameters \( \theta \)) and the observed values, \( \mathbf{H}_{\text{obs}} \). The covariance matrix \( \mathbf{C} \) accounts for the variances \( \sigma_H^2(z_i) \) of the measurements, assuming the dataset points are uncorrelated. Its inverse, \( \mathbf{C}^{-1} \), incorporates the uncertainties in the observations.
    \item Type Ia Supernovae: We consider the SNe Ia dataset without the SHOES calibration \cite{brout2022pantheon}. The chi-square statistic for this dataset is given by the following expression: $\chi^2_{\text{SNe Ia}} = \Delta \mathbf{D}^T \mathbf{C}^{-1}_{\text{total}} \Delta \mathbf{D},$ where \( \Delta \mathbf{D} \) represents the differences between the observed distance moduli, \( \mu(z_i) \), and the model predictions, \( \mu_{\text{model}}(z_i, \theta) \) at redshift $z_i$ ( calculated for model parameters $\theta$). The differences are calculated as: $\Delta D_i = \mu(z_i) - \mu_{\text{model}}(z_i, \theta).$ The total covariance matrix, \( \mathbf{C}_{\text{total}} \), combines both statistical uncertainties (\( \mathbf{C}_{\text{stat}} \)) and systematic uncertainties (\( \mathbf{C}_{\text{sys}} \)). Its inverse, \( \mathbf{C}^{-1}_{\text{total}} \), is used in the calculation. The model predicted distance modulus, \( \mu_{\text{model}}(z_i) \), is given by: $\mu_{\text{model}}(z_i) = 5 \log_{10} \left( \frac{d_L(z)}{\text{Mpc}} \right) + \mathcal{M} + 25,$ where \( d_L(z) \) is the luminosity distance in a flat FLRW Universe, expressed as: $d_L(z) = c(1 + z) \int_0^z \frac{dz'}{H(z')}.$ Here, \( c \) is the speed of light in vaccum, and \( H(z) \) is the Hubble parameter.\\\\
    \item Baryon Acoustic Oscillations: We use the Baryon Acoustic Oscillations (BAO) measurements from the Dark Energy Spectroscopic Instrument Year 1 (DESI Y1) \cite{adame2024desi}. These measurements rely on the sound horizon at baryon decoupling, which is approximately \( r_d \approx 147.09 \pm 0.26 \, \text{Mpc} \) \cite{planck2018results}. To ensure model independence approach, \( r_d \) is treated as a free parameter, allowing constraints on \( r_d \) without assumptions about the physics of recombination or the early Universe \cite{pogosian2020recombination,jedamzik2021reducing,pogosian2024consistency,lin2021early,vagnozzi2023seven}.To analyze the BAO dataset, we calculate the following key cosmological distances under the assumption of a flat Universe. The Hubble distance, \( D_H(z) = \frac{c}{H(z)} \), The comoving angular diameter distance, \( D_M(z) = c \int_0^z \frac{dz'}{H(z')} \), The volume-averaged distance, \( D_V(z) = \left[ z D_M^2(z) D_H(z) \right]^{1/3} \). Here, \( H(z) \) is the Hubble parameter, and \( c \) is the speed of light in vacuum. Consequently, we calculate ratios of these distances to the sound horizon, \( r_d \): $\frac{D_M(z)}{r_d}, \quad \frac{D_H(z)}{r_d}, \quad \frac{D_V(z)}{r_d}.$ The chi-square statistic for each ratio is defined as: $\chi^2_{D_Y / r_d} = \Delta D_Y^T \cdot \mathbf{C}^{-1}_{D_Y} \cdot \Delta D_Y,$ where \( \Delta D_Y = D_{Y / r_d, \text{Model}} - D_{Y / r_d, \text{Data}} \), with \( Y \) representing \( H, M, \) or \( V \). The covariance matrix \( \mathbf{C}_{D_Y} \), typically diagonal with variances \( \sigma_{D_Y}^2 \), is inverted to \( \mathbf{C}^{-1}_{D_Y} \). The total chi-square for the BAO data is then given by: $\chi^2_{\text{BAO}} = \chi^2_{D_H / r_d} + \chi^2_{D_V / r_d} + \chi^2_{D_M / r_d}.$
\end{itemize}
The total chi-squared statistic, \(\chi^2_{\text{Tot}}\), is then defined as the sum of contributions from all datasets:
\[
\chi^2_{\text{Tot}} = \chi^2_{\text{CC}} + \chi^2_{SNe Ia} + \chi^2_{BAO} 
\]
In our analysis, the Interacting Model have six free parameters, while the Non-Interacting Model have seven free. parameters along with one derived parameters, \( \Omega_{de_0} \). The distribution of the \( \Omega_{de_0} \) is extracted using the relation: 
$\Omega_{de_0} = 1 - \Omega_{dm_0} - \Omega_{b0}$. This approach eliminates the need to explicitly vary \( \Omega_{de_0} \), as it is fully determined by the other parameters in the model. For data visualization, we use the \texttt{GetDist} library \cite{lewis2019getdist}, which generates 1D and 2D posterior distribution plots to better analyze the parameter constraints.
\begin{table}
\begin{tabular}{|c|c|c|c|}
\hline
Models & Parameter & Prior & JOINT  \\
\hline
& $H_{0}$ & $[50.,100.]$ & $67.73{\pm 1.1}$  \\[0.1cm]
& $\Omega_{m0}$ &$[0.,1.]$   & $0.3275{\pm 0.0065}$  \\[0.1cm]
$\Lambda$CDM Model & $\Omega_{\Lambda0}$ &$[0.,1.]$   & $0.6724{\pm 0.0078}$  \\[0.1cm]
&$r_{d} (Mpc)$   &$[100.,300.]$  &$147.1{\pm 2.2}$  \\[0.1cm]
& $\mathcal{M}$ & $[-20.,-18.]$  &$-19.422{\pm 0.027}$  \\[0.1cm]
\hline
& $H_{0}$ & $[50.,100.]$ & $65.1{\pm 1.0}$  \\[0.1cm]
& $\Omega_{b0}$ &$[0.,0.1]$   &$0.04569{\pm 0.00041}$ \\[0.1cm]
& $\Omega_{dm_{0}}$ &$[0.,1.]$   &$0.2715{\pm 0.0090}$ \\[0.1cm]
Non-Interacting & $\beta$ &$[0.,1.]$   &$0.292{\pm 0.054}$ \\[0.1cm]
Model & $\Omega_{de_{0}}$ &$[0.,1.]$   &$0.6828{\pm 0.0090}$ \\[0.1cm]
&$r_{d} (Mpc)$   &$[100.,300.]$  &$146.6{\pm 2.3}$  \\[0.1cm]
& $\mathcal{M}$ & $[-20.,-18.]$  &$-19.427{\pm 0.034}$  \\[0.1cm]
\hline
& $H_{0}$ & $[50.,100.]$ & $65.0{\pm 1.1}$  \\[0.1cm]
& $\Omega_{b0}$ &$[0.,0.1]$   &$0.04534{\pm 0.00037}$ \\[0.1cm]
Interacting& $\Omega_{dm_{0}}$ &$[0.,1.]$   &$0.2235{\pm 0.0083}$ \\[0.1cm]
Model& $\beta$ &$[0.,1.]$   &$0.380_{- 0.044}^{+0.048}$ \\[0.1cm]
& $\gamma$ & $[-0.1,0.1]$   &$-0.0500{\pm 0.0036}$ \\[0.1cm]
& $\Omega_{de_{0}}$ &$[0.,1.]$   &$0.7312{\pm 0.0083}$ \\[0.1cm]
&$r_{d} (Mpc)$   &$[100.,300.]$  &$146.7{\pm 2.3}$  \\[0.1cm]
& $\mathcal{M}$ & $[-20.,-18.]$  &$-19.428{\pm 0.035}$  \\[0.1cm]
\hline
\end{tabular}
\caption{MCMC Results}\label{tab_1}
\end{table}
\begin{figure*}
\centering
\includegraphics[width=16.0 cm]{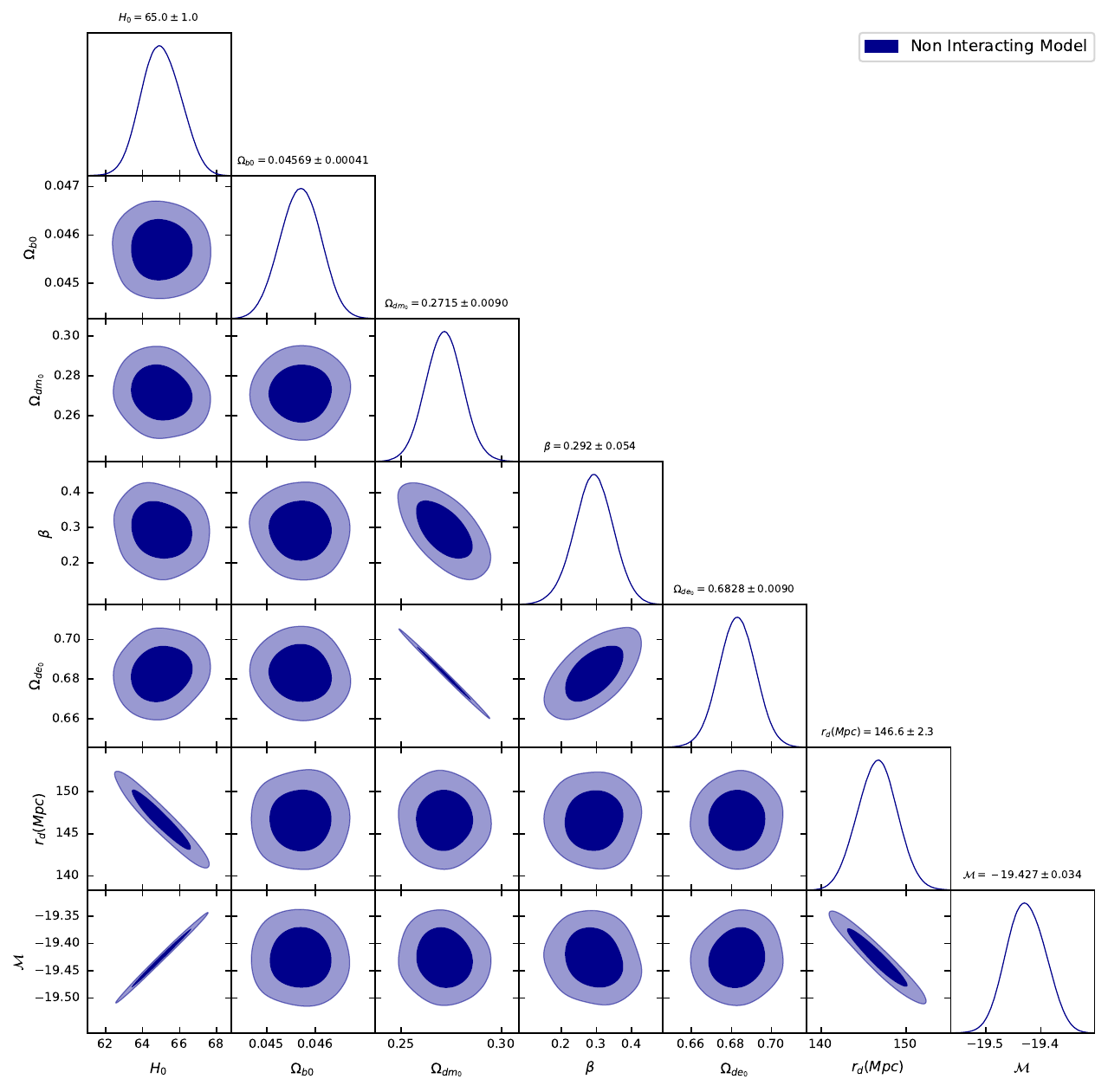}
\caption{The confidence contours at the 1\(\sigma\) and 2\(\sigma\) levels based on constraints for the linear model within the semi-symmetric gravity framework.}\label{fig_2}
\end{figure*}
\begin{figure*}
\centering
\includegraphics[width=18.0 cm]{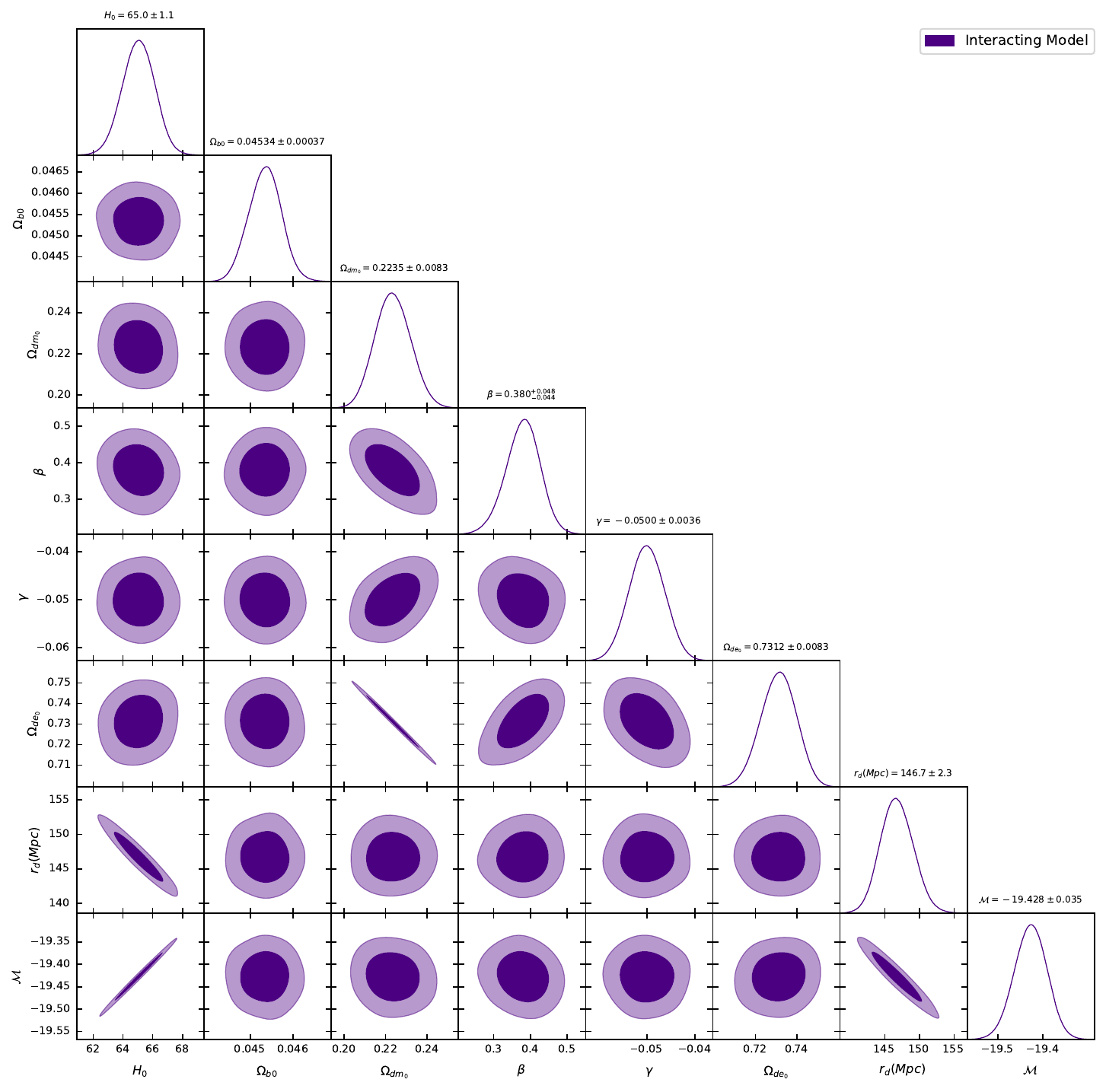}
\caption{The confidence contours at the 1\(\sigma\) and 2\(\sigma\) levels based on constraints for the linear model within the semi-symmetric gravity framework.}\label{fig_3}
\end{figure*}
\begin{figure*}
\centering
\includegraphics[scale=0.53]{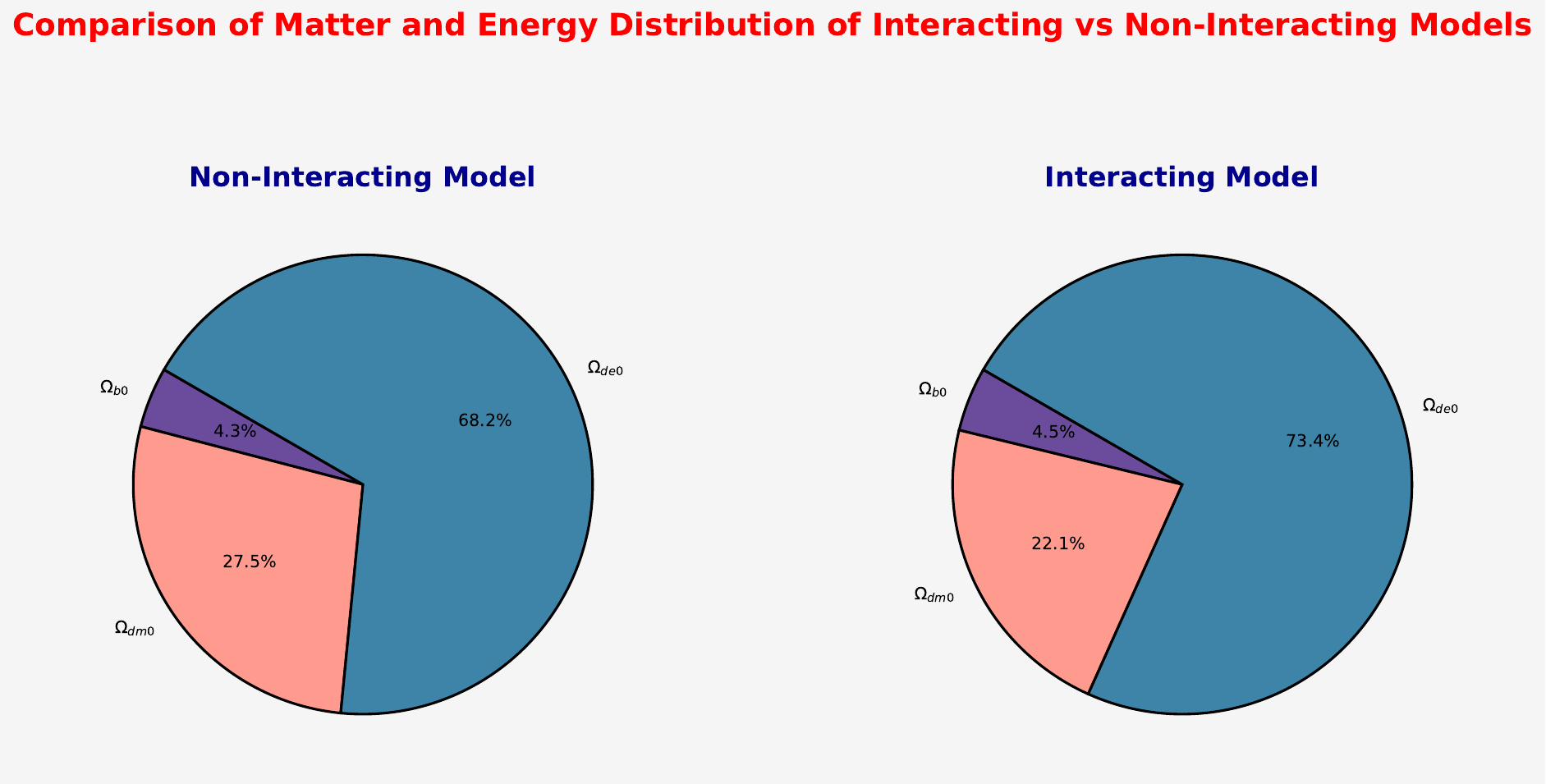}
\caption{Comparison of Matter and Energy Distribution: Interacting vs Non-Interacting Cosmological Models}\label{fig_4}
\end{figure*}
\subsection{Comparative Analysis with the Standard Model and Observational Data}
To evaluate the Interacting and Non-Interacting models in describing the Universe's expansion, we compare them with observational data and the standard cosmological model (\(\Lambda\)CDM). This analysis involves examining key parameters such as the Hubble parameter and distance modulus (\(\mu(z)\)) both as functions of redshift. Observational datasets from cosmic chronometers (CC) and Type Ia supernovae (SNe Ia) are used to assess the models' ability to match real-world data. This approach offers insights into how well these models align with observations and evaluates their potential as alternatives to \(\Lambda\)CDM for explaining the Universe's late-time expansion.
\subsubsection{Comparison of Models Using the Hubble Parameter and distance modulus}
We begin by analyzing the Hubble parameter \(H(z)\), which describes how the Universe's expansion rate changes over time. For the standard \(\Lambda\)CDM model, \(H(z)\) is expressed as: $H(z) = H_0 \sqrt{\Omega_{m0} (1 + z)^3 + \Omega_{\Lambda 0}}.$ Here, \(H_0 = 68.2 \, \text{km/s/Mpc}\), \(\Omega_{m0} = 0.311\), and \(\Omega_{\Lambda 0} = 0.686\). The evolution of \(H(z)\) as a function of redshift is plotted for the Interacting, Non-Interacting, and \(\Lambda\)CDM models and compared with the CC dataset. Further, we compute the distance modulus \(\mu(z)\) for the Interacting and Non-Interacting models, comparing them with \(\Lambda\)CDM. The distance modulus is given by \(\mu(z) = 5 \log_{10}(D_L(z)) + 25\), where \(D_L(z)\), the luminosity distance, is defined as \(D_L(z) = (1 + z) D_C(z)\). The comoving distance \(D_C(z)\) is computed as: $D_C(z) = \int_0^z \frac{c}{H(z')} \, dz',$  
where \(c\) is the speed of light in vacuum, and \(H(z')\) is the Hubble parameter at redshift \(z'\). Using the mean values of each parameters from the MCMC analysis, we calculate the distance modulus for the Interacting and Non-Interacting models (\(\mu_{\text{Model}}(z)\)) and compare them with \(\Lambda\)CDM (\(\mu_{\Lambda \text{CDM}}(z)\)). These theoretical predictions are plotted against observational data from 1701 Type Ia supernovae (SNe Ia).
\begin{figure*}[htb]
\begin{subfigure}{.48\textwidth}
\includegraphics[width=\linewidth]{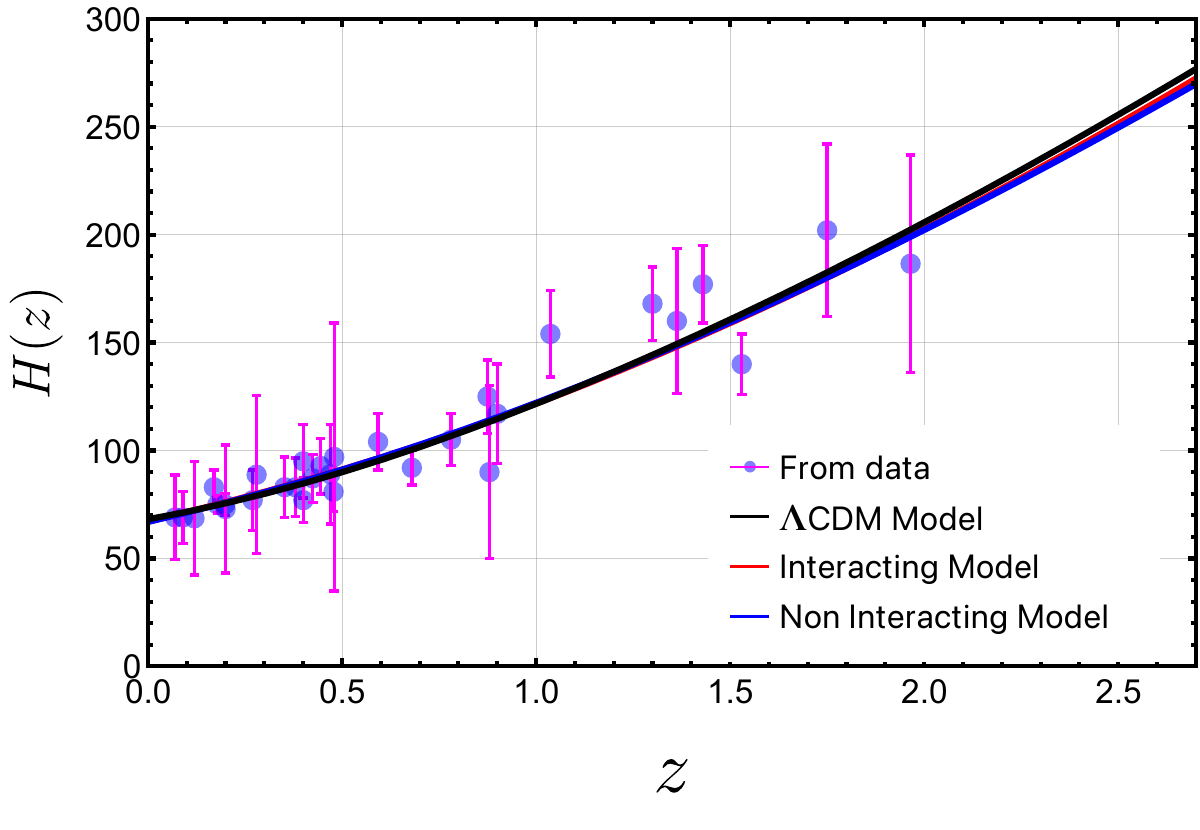}
    \caption{Hubble parameter \( H(z) \)}
    \label{fig_5a}
\end{subfigure}
\hfil
\begin{subfigure}{.48\textwidth}
\includegraphics[width=\linewidth]{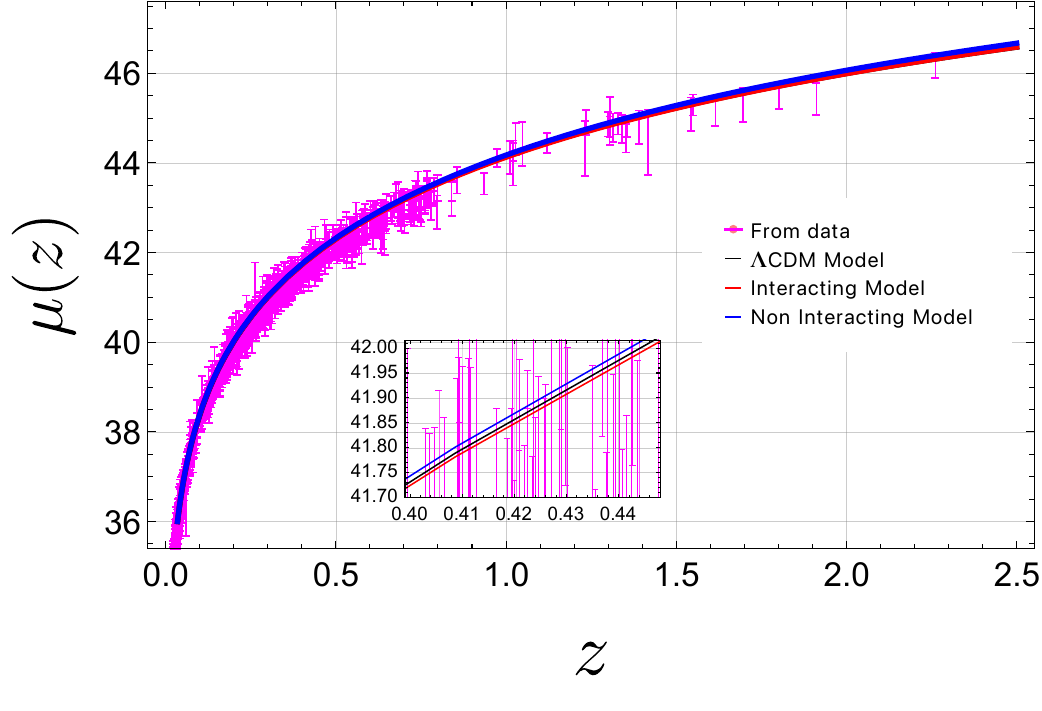}
    \caption{Distance Modulus \( \mu(z) \)}
    \label{fig_5b}
\end{subfigure}
\caption{The evolution of the Hubble parameter and Distance Modulus as a function of redshift is presented for the \(\Lambda\)CDM model (black line), the Interacting model (red line), and the Non-Interacting model (blue line). These are compared against the observational Hubble measurements, represented by purple dots with corresponding error bars shown as magenta lines.}\label{fig_5}
\end{figure*}
\section{Statistical Metrics}\label{sec4}
To distinguish the performance of the Interacting and Non-Interacting models from the \(\Lambda\)CDM framework, several statistical metrics were utilized. One essential metric for evaluating the goodness of fit is the reduced chi-squared (\(\chi^2_{\text{red}}\)), expressed as:  
\[
\chi^2_{\text{red}} = \frac{\chi_{\text{tot,min}}^2}{\text{DOF}},
\]  
where \(\chi_{\text{tot,min}}^2\) denotes the minimum total chi-squared value, and DOF (degrees of freedom) is calculated as the difference between the number of data points and the number of free parameters in the model. A value of \(\chi^2_{\text{red}} \approx 1\) signifies a model that fits well, \(\chi^2_{\text{red}} < 1\) suggests potential overfitting, and \(\chi^2_{\text{red}} \gg 1\) points to a poor fit \cite{andrae2010and}. To further compare the models, the Akaike Information Criterion (AIC) and the Bayesian Information Criterion (BIC) were computed \cite{liddle2007information,akaike1974new,burnham2004multimodel,schwarz1978estimating}. The AIC is determined as:  
\[
\mathrm{AIC} = \chi_{\text{tot,min}}^2 + 2k,
\]  
while the BIC is given by:  
\[
\mathrm{BIC} = \chi_{\text{tot,min}}^2 + k \ln N,
\]  
where \(k\) represents the number of model parameters, and \(N\) is the total number of data points. Both AIC and BIC assess model quality by considering the trade-off between fit and complexity, with smaller values indicating a more favorable model. To evaluate the performance of the Interacting and Non-Interacting models relative to \(\Lambda\)CDM, the differences in AIC and BIC values were calculated as:  
\[
\Delta \mathrm{AIC} = \mathrm{AIC}_{\text{Interacting/Non-Interacting}} - \mathrm{AIC}_{\Lambda\text{CDM}},
\]  
\[
\Delta \mathrm{BIC} = \mathrm{BIC}_{\text{Interacting/Non-Interacting}} - \mathrm{BIC}_{\Lambda\text{CDM}}.
\]  
Based on the Jeffreys scale \cite{jeffreys1998theory}, \(|\Delta \mathrm{AIC}| \leq 2\) implies no significant difference between models, whereas \(|\Delta \mathrm{AIC}| \geq 4\) indicates the model with the higher AIC is less favored. For BIC, \(|\Delta \mathrm{BIC}| \leq 2\) corresponds to weak evidence against a model, \(2 < |\Delta \mathrm{BIC}| \leq 6\) indicates strong evidence, and \(|\Delta \mathrm{BIC}| > 6\) suggests very strong evidence against the higher BIC model. Negative \(\Delta \mathrm{AIC}\) or \(\Delta \mathrm{BIC}\) values indicate preference for the Interacting or Non-Interacting models over \(\Lambda\)CDM. Additionally, the statistical significance of the fit was assessed using the p-value \cite{wasserstein2016asa}, calculated as:  
\[
p = 1 - F_{\chi_{\text{tot,min}}^2}(\chi \mid \nu),
\]  
where \(F_{\chi_{\text{tot,min}}^2}(\chi \mid \nu)\) represents the cumulative distribution function (CDF) of the chi-squared distribution, and \(\nu\) is the degrees of freedom. A p-value less than 0.05 (\(p < 0.05\)) indicates statistical significance, providing strong evidence to reject the null hypothesis.
\begin{table*}[htbp]
\begin{center}
\begin{tabular}{|c|c|c|c|c|c|c|c|c|c|}
\hline
Model & ${\chi_{\text{tot},min}^2}$ & $N_{tot}$ & $k$ & $\chi_{\text {red}}^2$ & AIC & $\Delta$AIC & BIC & $\Delta$BIC & p-value \\[0.1cm]
\hline
$\Lambda$CDM Model & 1782.14 & 1744 & 4 & 1.024 & 1790.14 & 0 & 1811.99 & 0 & 0.235 \\[0.1cm] 
\hline
Non-Interacting Model & 1768.95 & 1744 & 6 & 1.017 & 1780.95 & -9.19 & 1813.73 & 1.74 & 0.296 \\[0.1cm] 
\hline
Interacting Model & 1764.12 & 1744 & 7 & 1.015 & 1778.12 & -12.02 & 1816.36 & 4.37 & 0.319 \\[0.1cm] 
\hline
\end{tabular}
\caption{Statistical Metrics for $\Lambda$CDM, Interacting, and Non-Interacting models, including \(\chi_{\text{tot,min}}^2\), \(\chi^2_{\text{red}}\), AIC, BIC, \(\Delta\mathrm{AIC}\), \(\Delta\mathrm{BIC}\) and p-values}\label{tab_2}
\end{center}
\end{table*}
\section{Results}\label{sec5}
Fig~\ref{fig_1} shows the evolution of the deceleration parameter (DP) as a function of redshift for the Interacting and Non-Interacting Models compared to the $\Lambda$CDM Model. At high redshifts (\(z = 3\)), the Interacting Model predicts a higher deceleration parameter (DP) of 0.484 compared to the Non-Interacting Model (0.422) and the $\Lambda$CDM Model (0.451), indicating a slower expansion in the Interacting Model due to a reduced influence of dark energy. At the present epoch (\(z = 0\)), the $\Lambda$CDM Model predicts the most negative DP value $(-0.532)$, reflecting the fastest acceleration of the universe’s expansion. In contrast, the Non-Interacting Model predicts slower acceleration (\(DP = -0.365\)), while the Interacting Model (\(DP = -0.416\)) suggests moderate acceleration, with higher acceleration in the Interacting Model due to a greater dark energy composition compared to the Non-Interacting Model. The transition redshift, \(z_{\text{tr}}\), marking the shift from deceleration to acceleration, differs across models. The $\Lambda$CDM Model predicts \(z_{\text{tr}} = 0.644\), the Interacting Model predicts \(z_{\text{tr}} = 0.730\), and the Non-Interacting Model predicts \(z_{\text{tr}} = 0.694\). The earlier transition to accelerated expansion in the Interacting Model suggests a stronger interaction between dark energy and dark matter. These results highlight that the Non-Interacting Model implies a slower rate of cosmic acceleration, attributed to a higher dark energy content in the absence of interaction. Figs ~\ref{fig_2} and \ref{fig_3} show the posterior distribution and triangle plot of the parameters for the Non-Interacting and Interacting Models, highlighting the 1$\sigma$ (68.3\%) and 2$\sigma$ (95.4\%) confidence contours. The triangle plot displays the marginal distributions of each parameter along the diagonal, while the off-diagonal elements represent the 2D joint distributions between pairs of parameters. Table ~\ref{tab_1} shows the mean values of cosmological parameters for the $\Lambda$CDM, Non-Interacting, and Interacting Models. The $\Lambda$CDM model predicts a present-day Hubble constant (\(H_0 = 68.1\)), which is in agreement with the value predicted by the Planck Collaboration and aligns with the DESI Collaboration's prediction. The matter-energy densities in the $\Lambda$CDM model are consistent with the values predicted by Planck (\(\Omega_{m0} = 0.315 \pm 0.007\) and \(\Omega_{\Lambda 0} = 0.685 \pm 0.007\)), and the predicted value of the sound horizon, \(r_d\), is also close to the Planck Collaboration's estimate. The Non-Interacting Model predicts a baryonic density (\(\Omega_{b0} = 0.04569\)), indicating that 4.5\% of the Universe is composed of ordinary matter. In contrast, the Interacting Model predicts a slightly lower baryonic density (\(\Omega_{b0} = 0.04534\)), suggesting that 4.3\% of the Universe consists of ordinary matter. The Non-Interacting Model predicts a dark matter density of \(\Omega_{dm_0} = 0.2715\), while the Interacting Model predicts a lower value of \(\Omega_{dm_0} = 0.2235\). In contrast, the Non-Interacting Model predicts a dark energy density of \(\Omega_{de_0} = 0.6828\), while the Interacting Model predicts a higher value of \(\Omega_{de_0} = 0.7312\). This results in a higher overall dark energy composition in the Interacting Model, where it constitutes 73.4\% of the total energy density, compared to 68.2\% in the Non-Interacting Model. Furthermore, the dark matter content in the Interacting Model is 22.1\%, while the Non-Interacting Model predicts 27.5\% dark matter. The relative composition of ordinary matter, dark energy, and dark matter in both models is illustrated in Fig.~\ref{fig_4}. These results suggests that in the Interacting Model, the increased dark energy content, due to the interaction parameter \(\gamma\), leads to a reduction in the dark matter fraction. Additionally, the predicted value of the sound horizon, \( r_d \), in both cosmological models is in close agreement with the value predicted by the Planck Collaboration. Fig.~\ref{fig_5} shows the evolution of the Hubble parameter \( H(z) \) and the Distance Modulus \( \mu(z) \) as a function of redshift. Fig.~\ref{fig_5a} shows the evolution of the Hubble parameter for the \(\Lambda\)CDM, Interacting, and Non-Interacting Models, compared with observational Hubble measurements. At redshifts \( z > 2 \), both the Interacting and Non-Interacting Models show slight deviations from the \(\Lambda\)CDM Model, while at \( z < 2 \), all models align closely with each other. Fig.~\ref{fig_5b} shows the evolution of the Distance Modulus for the \(\Lambda\)CDM, Interacting, and Non-Interacting Models, compared with observational SNe Ia measurements. At higher redshifts (\( z > 1 \)), the Interacting and Non-Interacting Models exhibit noticeable deviations from the \(\Lambda\)CDM Model, but at lower redshifts (\( z < 1 \)), they closely match the \(\Lambda\)CDM predictions. These differences are subtle and are more evident in the subfigure in the corresponding plot. Table~\ref{tab_2} presents a comparison of the \(\Lambda\)CDM, Interacting, and Non-Interacting models based on various statistical metrics. The reduced chi-squared (\(\chi^2_{\text{red}}\)) values are used to assess the goodness of fit. A value close to 1 indicates that the model fits the data well, while higher values suggest a poorer fit. The \(\Lambda\)CDM model has a \(\chi^2_{\text{red}} = 1.024\), meaning it fits the data reasonably well with a good balance between fit and complexity. The Non-Interacting model has a similar \(\chi^2_{\text{red}} = 1.017\), showing a very slight improvement over the \(\Lambda\)CDM model. The Interacting model's \(\chi^2_{\text{red}} = 1.015\) is almost identical to that of the Non-Interacting model, indicating that adding interacting parameter doesn’t significantly improve the model's fit compared to the Non-Interacting model. The AIC and BIC are other important metrics used to evaluate models. These criteria consider both the fit and the complexity of a model. The \(\Lambda\)CDM model has an AIC of 1790.14 and a BIC of 1811.99, which are used as reference values for comparison. The Non-Interacting model has slightly lower values for both AIC (1780.95) and BIC (1813.73), suggesting a small improvement over the \(\Lambda\)CDM model. However, the differences are very small (\(\Delta \text{AIC} = -9.19\), \(\Delta \text{BIC} = 1.74\)), indicating only weak evidence in favor of the Non-Interacting model. The Interacting model has an even lower AIC of 1778.12, but its BIC (1816.36) is higher than that of the \(\Lambda\)CDM model, implying that the Interacting model is slightly more complex and doesn’t offer a significantly better fit. The \(\Delta \text{AIC} = -12.02\) and \(\Delta \text{BIC} = 4.37\) values suggest that while the Interacting model has a marginally better AIC, the higher BIC indicates that the simpler \(\Lambda\)CDM model might be preferred. Consequently , the p-values for all models are above 0.05: \(\Lambda\)CDM (0.235), Non-Interacting (0.296), and Interacting (0.319). This means that none of the models provide strong evidence to reject the null hypothesis, suggesting that the differences between the models are not statistically significant.
\section{Conclusion} \label{sec6}
This study provides a thorough comparative analysis of interacting and non-interacting dark energy models within the FLRW framework. By leveraging the latest observational datasets, including BAO measurements from DESI Year 1, cosmic chronometers, and the Pantheon\(^{+}\) compilation, we determine the mean values of matter and energy densities for both frameworks. A key feature of our approach is treating \(r_d\) as a free parameter, allowing us to adopt a model-independent perspective that minimizes assumptions about early Universe physics. This enables late-time cosmological observations to directly constrain the value of \(r_d\). Our analysis reveals that interacting models suggest a higher contribution of dark energy (\(73.4\%\)) to the total cosmic energy density compared to non-interacting models (\(68.2\%\)). This underscores the potential significance of the interaction parameter in understanding the dynamics of dark energy and its relationship with dark matter. To contextualize these findings, we compare the \(\Lambda\)CDM paradigm with both interacting and non-interacting dark energy models. A detailed examination of the deceleration parameter (DP) highlights differences in cosmic evolution between the models. At high redshifts (\(z = 3\)), the interacting model predicts a slower expansion due to a reduced influence of dark energy. In contrast, the \(\Lambda\)CDM model exhibits the fastest acceleration at the present epoch (\(z = 0\)). The transition to accelerated expansion occurs earlier in the interacting model, indicating a stronger interaction between dark energy and dark matter. Additionally, the interacting model predicts a higher dark energy content and slightly lower dark matter density compared to the non-interacting model, a direct result of the interaction parameter. Regarding cosmological evolution, both the Hubble parameter and distance modulus show small deviations between the interacting and non-interacting models when compared to \(\Lambda\)CDM predictions, particularly at higher redshifts. These differences diminish at lower redshifts, reflecting the dominance of dark energy in all models at later times. Statistical metrics, including reduced chi-squared (\(\chi^2_{\text{red}}\)), AIC, \(\Delta\)AIC, BIC, \(\Delta\)BIC, and p-values, reveal that all three models \(\Lambda\)CDM, interacting, and non interacting provide a good fit to the data. The reduced chi-squared values are comparable, indicating that each model successfully captures the main features of cosmic expansion. While the non-interacting model demonstrates a slight improvement over the \(\Lambda\)CDM model, the interacting model performs similarly to the non-interacting model but with a slightly higher BIC, suggesting that the added complexity of interaction effects is not strongly favored by the data. Furthermore, the p-values for all models exceed 0.05, indicating that the observed differences between the models are not statistically significant.Interacting dark energy models are motivated by the unresolved nature of dark energy and dark matter, which constitute the majority of the energy density of the Universe. While \(\Lambda\)CDM assumes these components evolve independently, interacting models explore the possibility of energy exchange between dark matter and dark energy. This interaction could address persistent issues in cosmology, such as the coincidence problem, which questions why dark energy and dark matter densities are of the same order of magnitude today despite their distinct evolutionary histories. Interacting models also offer alternative explanations for cosmic acceleration and deviations in growth rates of large-scale structure, potentially bridging the gap between theoretical predictions and observations. By examining the potential interaction between dark energy and dark matter, we aim to deepen our understanding of these elusive components, refine cosmological models, and provide insights into the fundamental mechanisms governing the evolution of the Universe.
\bibliographystyle{elsarticle-num}
\bibliography{mybib}
\end{document}